\documentclass[sigconf]{acmart}

\usepackage{booktabs} 
\usepackage{pifont}
\usepackage{color}
\usepackage{adjustbox}

\newcommand{\RI}{RouterInfo}
\newcommand{\LS}{LeaseSet}
\newcommand{\ES}{eepsite}

\begin{document}
\title[Measuring the I2P Anonymity Network and its Censorship Resistance]{An Empirical Study of the I2P Anonymity Network and its Censorship Resistance}

\author{Nguyen Phong Hoang}
\affiliation{%
  \institution{Stony Brook University}
  \city{Stony Brook}
  \state{New York}
}
\email{nghoang@cs.stonybrook.edu}

\author{Panagiotis Kintis}
\affiliation{%
  \institution{Georgia Institute of Technology}
  \city{Atlanta}
  \state{Georgia}
}
\email{kintis@gatech.edu}

\author{Manos Antonakakis}
\affiliation{%
  \institution{Georgia Institute of Technology}
  \city{Atlanta}
  \state{Georgia}
}
\email{manos@gatech.edu}

\author{Michalis Polychronakis}
\affiliation{%
  \institution{Stony Brook University}
  \city{Stony Brook}
  \state{New York}
}
\email{mikepo@cs.stonybrook.edu}

\renewcommand{\shortauthors}{NP. Hoang et al.}

\copyrightyear{2018} 
\acmYear{2018} 
\setcopyright{acmcopyright}
\acmConference[IMC '18]{2018 Internet Measurement Conference}{October 31-November 2, 2018}{Boston, MA, USA}
\acmBooktitle{2018 Internet Measurement Conference (IMC '18), October 31-November 2, 2018, Boston, MA, USA}
\acmPrice{15.00}
\acmDOI{10.1145/3278532.3278565}
\acmISBN{978-1-4503-5619-0/18/10}

\begin{abstract}{Tor and I2P are well-known anonymity networks used by many
individuals to protect their online privacy and anonymity. Tor's centralized
directory services facilitate the understanding of the Tor network, as well as
the measurement and visualization of its structure through the Tor Metrics
project. In contrast, I2P does not rely on centralized directory servers, and
thus obtaining a complete view of the network is challenging. In this work, we
conduct an empirical study of the I2P network, in which we measure properties including
population, churn rate, router type, and the geographic distribution of I2P
peers. We find that there are currently around 32K active I2P peers in the
network on a daily basis. Of these peers, 14K are located
behind NAT or firewalls.

Using the collected network data, we examine the blocking resistance
of I2P against a censor that wants to prevent access to I2P using address-based
blocking techniques. Despite the decentralized characteristics of I2P, we
discover that a censor can block more than 95\% of peer IP addresses known by a
stable I2P client by operating only 10 routers in the network. This amounts to
severe network impairment: a blocking rate of more than 70\% is enough to cause
significant latency in web browsing activities, while blocking more than 90\% of
peer IP addresses can make the network unusable.
Finally, we discuss the security consequences of the network being blocked,
and directions for potential approaches to make I2P more resistant to
blocking.}\end{abstract}

\ccsdesc[500]{Networks~Network measurement}
\ccsdesc[500]{Networks~Network privacy and anonymity}

\keywords{I2P anonymity network, network metrics, Internet censorship,
blocking resistance}

\maketitle

\section{Introduction}
\label{sec:introduction}

In recent years, Internet censorship and surveillance have become
prevalent~\cite{Winter2012, Ensafi2015, Bloomberg2017, FreedomHouse2017,
  IranTelegram, RussiaTelegram}. For this reason, anonymous communication has
drawn attention from both researchers and Internet users~\cite{Tor04,
  petcon2009-zzz, Winter2012, FacebookTor2014, Nobori2014, Ensafi2015,
  Zamani2017}. As anonymous communication networks grow to support more users,
more anonymity and censorship circumvention tools are becoming freely
available~\cite{Hoang2014Anonymous}.
Some of these tools include proxy servers, Virtual Private
Network (VPN) services, the Onion Router (Tor)~\cite{Tor04}, and the Invisible
Internet Project (I2P)~\cite{petcon2009-zzz}. Tor and I2P are the most popular
low-latency anonymous communication networks, which use the onion routing
technique~\cite{syverson97anonymous} to protect user anonymity.

Although both Tor and I2P provide similar features, there are some major
differences between them. Tor operates at the TCP stream level, while I2P
traffic can use both TCP and UDP. Tor has a centralized architecture in which
a set of directory authorities keep track of the network, while no entity has
a complete view of the I2P network due to its decentralized nature. Every
I2P peer helps other peers to route traffic by default, while there are only
6.5K Tor routers serving more than two million users per day, as of May
2018~\cite{TorMetrics}. As a result, while Tor is mainly designed for
latency-sensitive activities (e.g., web browsing) due to bandwidth
scarcity~\cite{McCoy2008}, I2P's capacity also enables
bandwidth-intensive peer-to-peer
(P2P) applications (e.g., BitTorrent)~\cite{Timpanaro2011}. 

While helping users to browse the Internet anonymously, these networks also
provide hidden services (comprising the ``dark web'') in which the anonymity of
both senders and receivers is preserved, thus protecting their privacy. Because
of its popularity and the support of volunteer-based ``exit nodes'' to the
normal Internet, Tor has been widely used and extensively researched. On the
other hand, I2P has not been studied as comprehensively. We identify two
potential reasons I2P has been less appealing than Tor. First, I2P's purely
distributed network architecture, which lacks any centralized directory service,
makes it harder to measure. Second, the intermittent availability of exit nodes
causes I2P to operate as a self-contained network (which only serves hidden
services) most of the time, making it less attractive to users who want to
casually browse websites on the public Internet.

In this work, we aim to fill this research gap by conducting an empirical
measurement of the I2P network, which may help popularize I2P to both academic
researchers and Internet users, and contribute to understanding its structure
and properties. With those two goals in mind, our investigation aims to answer
the following main questions.

\emph{What is the population of I2P peers in the network?} While Tor relies on
a centralized architecture for tracking its public relays, which are indexed by
a set of hard-coded authority servers, I2P is a distributed P2P
network in which no single centralized authority can keep track of all active
peers~\cite{Grigg2017, Shahbar2017, Ali2016, Tchabe2014, Conrad2014,
  Zantout2011}. Tor developers can easily collect information about
the network and even visualize it, as part of the Tor Metrics
project~\cite{Wecsr10measuring-tor}. In contrast, there have been very few
studies attempting to measure the I2P network~\cite{Timpanaro2011, Liu2014, Gao2017}.

In this work, we attempt to estimate the size of the I2P network by running up to 40
I2P nodes under different settings for network monitoring purposes. We find that there
are currently 32K active I2P peers in the I2P network on a daily basis. The United
States, Russia, England, France, Canada, and Australia contribute more than 40\%
of these peers. Different from prior works, we also observed about 6K peers
that are from 30 countries with poor Press Freedom
scores~\cite{worldPressFreedom}. This is an indication that I2P is possibly being used as
an alternative to Tor in regions with heavy Internet censorship and surveillance. 

\emph{How resilient is I2P against censorship, and what is the cost of
  blocking I2P?} Despite the existence of many pro-privacy and anti-censorship
tools, these are often easily blocked by local Internet authorities, thus
becoming inaccessible or difficult to access by non-tech-savvy
users~\cite{dutton2011freedom}. Hence, it is important to not only develop
censorship-resistant communication tools, but also to ensure that they are
easily accessible to end users. Due to the centralized nature of Tor's network
architecture, it is relatively easy for a censor to obtain a list of all public
Tor routers and block them~\cite{ChinaTorBlock}. Even hidden routers (also
known as ``bridges'')
are often discovered and blocked~\cite{Ensafi2015, Arun2018}. Despite its
decentralized design, there have still been reported attempts to block
I2P~\cite{ChinaBlockade}. However, to the best of our knowledge, no prior
studies have analyzed how challenging (or not) it is for a censor to block I2P
access. By analyzing the data we collected about the I2P the network, we
examine the censorship resistance of I2P using a probabilistic model. We
discover that a censor can block more than 95\% of peer IP addresses known to a
stable I2P client by injecting only 10 routers into the network.

In summary, the primary contribution of this work is an empirical measurement of
the I2P network, that aims to not only improve our understanding of I2P's
network properties, but also to assess the vulnerability of the I2P network to
address-based blocking.

The rest of the paper is organized as follows. Section~\ref{sec:relatedWork}
gives an overall background of I2P
and presents related works.
As an indispensable part of an anonymity network study,
ethical considerations are discussed in Section~\ref{sec:ethics}, where we
justify the principles to which we adhere while collecting and analyzing data
for this study. In Section~\ref{sec:methodology}, we explain our measurement
methodology, including machine specifications, network bandwidths, and the I2P
router types that we used to conduct our measurements. The measurement results
(e.g., the population of I2P peers, churn rate, and peer distribution)
of the I2P network properties are analyzed in
Section~\ref{sec:NetworkMeasurement}. Based on these network properties, we then
examine the blocking resistance of the network in
Section~\ref{sec:blockingResistance}, where we discover that I2P is highly
vulnerable to address-based blocking in spite of its decentralized
nature. Finally, in Sections~\ref{sec:discussion}
and~\ref{sec:conclusion}, we conclude by discussing consequences of the
network being censored and introducing potential approaches to hinder I2P censorship
attempts using address-based blocking, based on the insights that we gained
from our network measurements.

\section{Background and Related Work}
\label{sec:relatedWork}

\subsection{I2P: The Invisible Internet Project}
\label{sec:i2p}

\subsubsection{Routing Mechanism}

\begin{figure}[t]
\centering
\includegraphics[width=1\columnwidth]{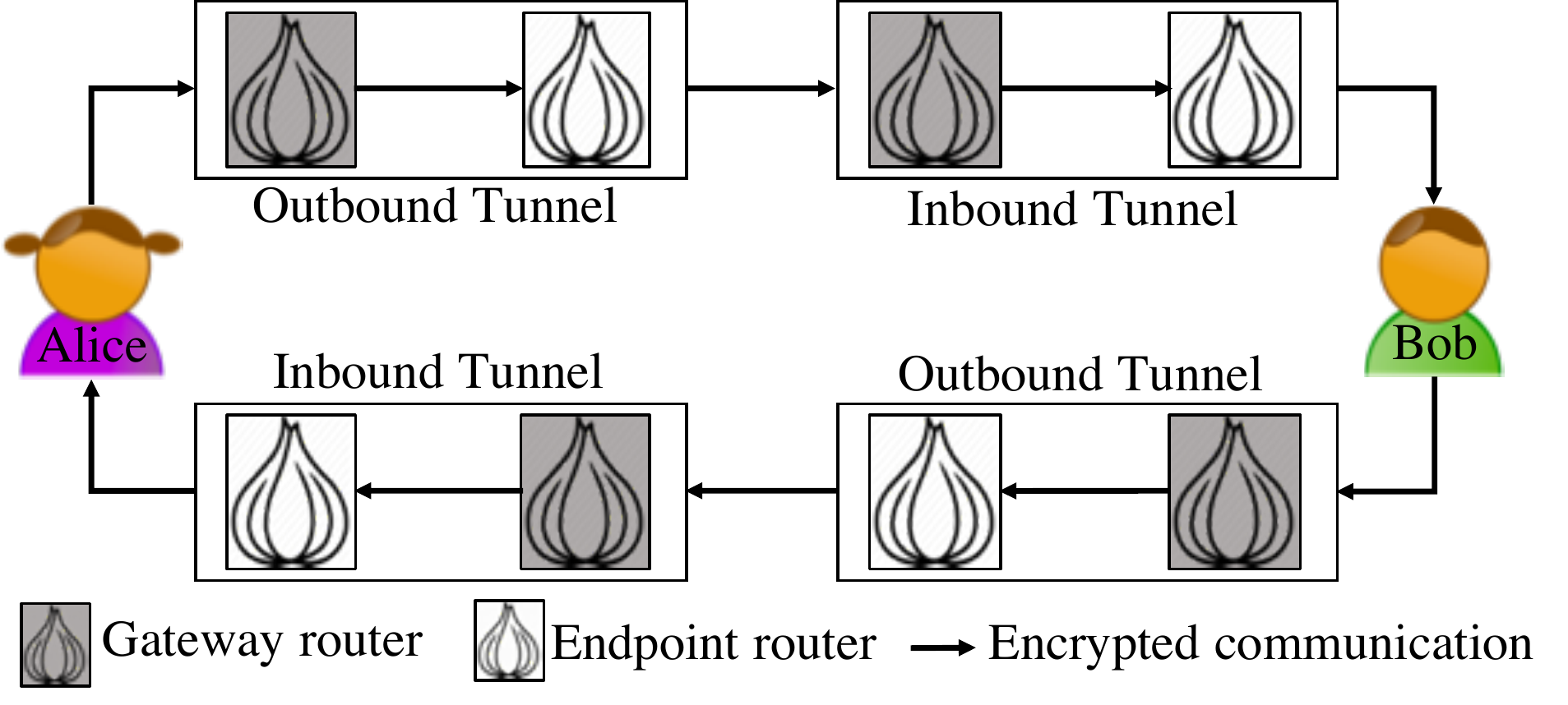}
\caption{Basic communication between two I2P peers using unidirectional tunnels~\protect\cite{architecture}.}
\label{fig:i2p_architecture}
\end{figure}

The Invisible Internet Project (I2P)~\cite{petcon2009-zzz}
is a message-oriented anonymous relay network consisting of peers (also
referred to as nodes, relays, or routers) running the I2P router software,
allowing them to communicate with each other. While Tor~\cite{Tor04} uses
onion-routing-based~\cite{syverson96OnionRouting, syverson97anonymous}
bidirectional circuits for communication, I2P utilizes
garlic-routing-based~\cite{Dingledine2000, freedman2000design, Dingledine2001}
unidirectional tunnels for incoming
and outgoing messages. An I2P client uses two types of communication tunnels:
inbound and outbound. Therefore, a single round-trip request message and its
response between two parties needs four tunnels, as shown in
Figure~\ref{fig:i2p_architecture}. For simplicity, each tunnel is depicted
with two hops. In practice, depending on the desired level of anonymity,
tunnels can be configured to comprise up to seven
hops~\cite{i2pTunnelRouting}. New tunnels are formed every ten minutes.

When Alice wants to communicate with Bob, she sends out messages on her outbound
tunnel. These messages head toward the gateway router of Bob's inbound tunnel.
Alice learns the address of Bob's gateway router by querying
a distributed network database~\cite{netDb} (discussed in more detail
in Section~\ref{sec:DHT}).
To reply to Alice, Bob follows the same process by
sending out reply messages on his outbound tunnel towards the gateway of
Alice's inbound tunnel. The anonymity of both Alice and Bob is preserved since
they only know the addresses of the gateways, but not each
other's real addresses. Note that gateways of inbound tunnels are published, while
gateways of outbound tunnels are known only by the party who is using them.

The example in
Figure~\ref{fig:i2p_architecture} illustrates a case in which I2P is used as a
self-contained network, with participating peers communicating solely with
each other. However, if Bob also provides an \emph{outproxy} service, Alice can
relay her traffic through Bob to connect to the public Internet.
The returned Internet traffic
is then securely relayed back to Alice by Bob via his outbound
tunnels, while Alice's identity remains unknown to both Bob and the visited
destination on the Internet.

Similar to Tor's onion routing, when an I2P message is sent over a tunnel
(i.e., from the gateway to the endpoint of that tunnel), it is encrypted
several times by the originator using the selected hops' public keys. Each hop
peels off one encryption layer to learn the address of the next hop where the
message needs to be forwarded to. When the message passes through an
inter-tunnel (i.e., from an outbound tunnel to an inbound tunnel), garlic
encryption (i.e. ElGamal/AES) is employed by the 
originator~\cite{i2pGarlicRouting}, adding an additional layer of end-to-end
encryption to conceal the message from the outbound tunnel endpoint and the
inbound tunnel gateway~\cite{architecture}.

Unlike Tor, multiple messages can be bundled together in a single I2P garlic
message. When they are revealed at the endpoint of the transmission tunnel,
each message, called "bulb"~\cite{freedman2000design} (or "clove" in I2P's
terminology~\cite{i2pGarlicRouting}), has its own delivery
instructions. Another major difference between Tor and I2P is that
all I2P nodes (except hidden routers, discussed in
Section~\ref{sec:i2pRouterPopulation}) also participate in the network as
relays, routing traffic for other nodes. In Figure~\ref{fig:i2p_architecture},
the hops (denoted by boxed onions) forming the tunnels for Alice and Bob
correspond to actual I2P users. While routing messages for Alice and Bob,
these hops can also communicate with their intended destinations in the same
way Alice and Bob do. Similarly, Alice and Bob can be chosen by other peers to
participate in the tunnels these peers will form.

\subsubsection{Distributed Directory}
\label{sec:DHT}

The network database of I2P, called \emph{netDb}, plays a vital role in the I2P
network by allowing peers to query for information about other peers and hidden
services. The network database is implemented as a distributed hash table using
a variation of the Kademlia algorithm~\cite{Maymounkov2002}. A newly joining peer
initially learns a small portion of the netDb through a bootstrapping process,
by fetching information about other peers in the network from a set of hardcoded
\emph{reseed} servers. Unlike Tor directory authorities, these reseed servers do
not have a complete view of the whole I2P network. They are equivalent to any
other peer in the network, with the extra ability to announce a small
portion of known routers to newly joining peers.

Queries for the network database are answered by a group of special
\emph{floodfill} routers~\cite{netDb}, which play an essential role in
maintaining the netDb. One of their main responsibilities is to store
information about peers and hidden services in the network in a decentralized
fashion using indexing keys (i.e. routing keys). These keys are calculated by a
SHA256 hash function of a 32-byte binary search key which is concatenated with a
UTC date string. As a result, these hash values change every day at UTC
00:00~\cite{netDb}. In the current I2P design, there are two ways to become a
floodfill router. The first option is to manually enable the floodfill mode from
the I2P router console. The other possibility is that a high-bandwidth router
could become a floodfill router automatically after passing several ``health''
tests, such as stability and uptime in the network, outbound message queue
throughput, delay, and so on.

The netDb contains two types of network metadata: \emph{\LS{}s} and
\emph{\RI{}s}. For instance, Bob's \LS{} tells Alice the contact information of
the tunnel gateway of Bob's inbound tunnel. A \RI{} provides contact information
about a particular I2P peer, including its key, capacity, address, and port. To
publish his \LS{}s, Bob sends a \emph{DatabaseStoreMessage} (DSM) message to
several floodfill routers, which encapsulates his \LS{}s. To query Bob's \LS{}
information, Alice sends a \emph{DatabaseLookupMessage} (DLM) to those floodfill
routers.

\subsection{Related Work}
\label{sec:relwork}

\subsubsection{I2P Network Measurement}

There have been only a few studies on monitoring I2P prior to this work. In 2011,
Timpanaro et al.~\cite{Timpanaro2011} built their monitoring architecture on
the Planet Lab testbed to characterize the usage of the I2P network. Planet
Lab is a network consisting of voluntary nodes run by research institutes and
universities around the globe. Therefore, bandwidth and traffic policies of
nodes running on this network are often restricted. As acknowledged by the
group, only 15 floodfill routers could be set up successfully due to the
bandwidth rate restrictions of Planet Lab, thus limiting the amount of
collected data. The authors later expanded their work to characterize the
usage of I2P, particularly the use of file-sharing applications in the
network~\cite{Timpanaro2012, Timpanaro2014}.

In 2014, Liu et al.~\cite{Liu2014} reported that they could observe 25,640
peers per day over a period of two weeks using various methods to discover
the network topology. However, there are some issues with the
methodology that the authors used to collect \RI{}s, which we will discuss in
later sections. More recently, Jeong et al.~\cite{Jeong2016} reported
leakage of .i2p domain name resolution queries in the public DNS
infrastructure. Russia, the USA, and China are top countries of leakage
sources. Gao et al.~\cite{Gao2017} conducted a study on the popularity and
availability of \emph{\ES{}s} (I2P's terminology for anonymous websites).
The authors claimed the discovery of 1,861 online \ES{}s, which made up over
80\% all anonymous websites in the I2P network.

\subsubsection{Anonymous Communication Network Blockage}

To the best of our knowledge, there has been no prior work focusing on the
blocking resistance of I2P. Throughout this paper, we aim to shed some light
on this aspect of the network. Similar to Tor or any other anonymous network,
I2P is susceptible to blockage. Prior to this study, there have been some
commercial tools alleging to be able to block I2P. However, to the best of our
knowledge, despite the range of techniques used by these tools, none are able
to block I2P effectively, or at least not to the degree that would be required
for a large-scale adoption (e.g., nationwide blocking). We briefly review some
of these tools below.

In network management, firewall rules are often employed to allow or filter
out traffic. Popular blocking techniques often base on port number, protocol
signature, and IP address. However, anonymity networks, including Tor and I2P,
are designed to withstand censorship~\cite{i2pCensorshipResistance,
theGuardianTor, torBridges}. As a result, any attempts to block these networks
could cause considerable collateral damage.

For port-based censorship, blocking onion relay ports (\emph{orports}) or
directory information exchange ports (\emph{dirports}) is effective enough to
block Tor relays, and blocking UDP port 123 would prevent I2P from functioning
properly because the I2P router software needs the Network Time Protocol (NTP)
service to operate properly. Nevertheless, many Tor relays have orports and
dirports running over port 80 (HTTP) or 443 (HTTPS), while many legitimate
applications also use port 123 for the NTP service. Furthermore, I2P is a P2P
network application that can run on a wide range of ports using both UDP and
TCP. More specifically, I2P can run on any arbitrary port in the range of
9000--31000~\cite{i2pPort}. As a result, port blocking is not ideal for
large-scale censorship because it can unintentionally block
the traffic of other legitimate applications.

As nationwide Internet censorship is growing worldwide, Deep Packet Inspection
(DPI) is widely used by various entities to detect the traffic pattern of
connections to anonymity networks~\cite{xu2016GFW, Choffnes2017AnEE, Li2017}.
Regardless of the use of well-known ports (i.e., 80, 443), the traffic of
connections to Tor entry relays is fingerprintable and easily blocked by DPI-
enabled firewall. Consequently, Tor's pluggable transports have been introduced to
cope with this problem~\cite{torPluggable}. These pluggable transports make
traffic from a client to Tor bridges look similar to other innocuous and
widely-used traffic. Similarly, the design of I2P also obfuscates its traffic
to prevent \emph{payload-analysis-based} protocol identification. However,
\emph{flow analysis} can still be used to fingerprint I2P traffic in the
current design because the first four handshake messages between I2P routers
can be detected due to their fixed lengths of 288, 304, 448, and 48
bytes~\cite{i2pNTCP1}. To solve this problem, the I2P team is working on the
development of an authenticated key agreement protocol that resists various
forms of automated identification and other attacks~\cite{i2pNTCP2}.

Tenable, a network security company, provides a firewall service that contains
some modules to detect I2P traffic. Based on our review of their guidelines,
none of them seem to be efficient in blocking I2P. For instance, one of the
guidelines for detecting I2P outbound traffic is to manually inspect the
system for any rogue process~\cite{TenableNetworkSecurity}, which may not be
feasible for large-scale blocking such as nationwide censorship.

SonicWALL, a company specialized in content control and network security,
suggests blocking I2P by filtering out both UDP and TCP tunnel traffic to
block proxy access with their App Control~\cite{SonicWALL}. However,
this approach is not feasible at a large scale either, as the company
acknowledges that the approach may cause collateral damage by unintentionally
blocking other legitimate traffic, such as encrypted UDP, IPSec VPN, and other
encrypted TCP traffic.

A more effective approach is destination filtering. To implement this
approach, a censor has to compile a list of active I2P peer addresses
and block access to all of them. This address-based blocking approach will
have a severe impact on the process of forming new I2P tunnels, thus
preventing users from accessing the I2P network. Furthermore, a simpler but
still effective way to prevent new users from accessing I2P is to block access to
I2P reseed servers, which are required for the bootstrapping process.
Consequently, first-time users will not be able to access the I2P network if
they are not able to fetch \RI{}s of other peers.\footnote{To cope with this
problem, I2P has a method for ``manual'' reseeding of a router, which we
discuss in Section~\ref{sec:reseedBlocking}.} One of the goals of our work is
to evaluate the cost and the effectiveness of the address-based
blocking approach against I2P.

\section{Ethical Considerations}
\label{sec:ethics}

Conducting research on anonymity networks comprising thousands of users must
be performed in a responsible manner that both respects user privacy, and does not
disrupt the operation of the network. It also necessitates all collected data
to be handled in a careful manner~\cite{Sicker2007}. Although I2P routers are run
by individuals who may actively use the I2P network for their own purposes,
our study does not involve any human subjects research, as it focuses on
studying the \emph{infrastructure} provided by I2P. Our measurements do not
capture users' traffic or their online activities. We solely measure
network-level characteristics of the I2P network.

To conduct our measurements, we need to introduce and operate several
additional routers into the live I2P network.
This is a standard approach in the context of studying anonymity
networks, as is evident by the many previous works that have followed it to study
the Tor network~\cite{McCoy2008,Biryukov2013,Biryukov2014,Sun2015,Singh2017}.
The I2P team also operates an I2P router to gather network information for
development purposes~\cite{petcon2009-zzz, i2pstats}. In particular, the
\texttt{stats.i2p} website provides network performance graphs to help the I2P
developers with monitoring the network and assessing the effectiveness
of software changes in each release.

The I2P community has come up with a set of
guidelines~\cite{I2PResearchGuidelines} for responsibly conducting research in
the I2P network, to which we strictly adhered. According to these guidelines,
we were in close contact with the I2P team regarding the purposes of our study
and our measurements. Adhering to the principle of minimizing the
collected data to only the absolutely necessary,
we collect from I2P's netDb
only each node's IP address, hash value, and capacity
information available in \RI{}s. Finally, we securely delete all
collected data after statistically analyzing them.
Only aggregated statistics about the collected data are published.

One could consider the (temporary) collection of IP addresses as a potential
violation of user privacy.
The topic of whether IP addresses are Personally Identifiable
Information (PII) is controversial across many jurisdictions~\cite{lah2008ip}.
As stated in Section 3.3.3 of the Guide to
Protecting the Confidentiality of Personally Identifiable Information
published by NIST~\cite{nistPII}, IP address not readily linkable
to databases or other sources that would identify specific individuals, are not
considered as PII. Therefore, the IP addresses observed in our measurements
cannot be considered PII, since they are not linkable to any other data
collected throughout our experiments that could be used to identifying
any individuals.
Note that the current design of I2P does not hide the use of
I2P from a user's Internet service provider (ISP)---the I2P router software only
helps to maintain the secrecy of messages and the anonymity between peers.
Nevertheless, we still need to analyze IP-related
data in a responsible manner that will minimize the risk of exposure to third
parties (before it is deleted). For instance, when mapping IP addresses to their
geographic location, we do not query any public APIs. Instead, we use a
locally installed version of the MaxMind Database to map them in an offline
fashion.

While previous works intensively crawled reseed servers and floodfill
routers to harvest the netDb~\cite{Liu2014}, we only monitor the
network in a \emph{passive} manner to avoid causing any interference or
unnecessarily overloading any I2P peers. I2P can be launched in a
virtual network mode for studies related to testing attacks on the
network~\cite{I2PResearchGuidelines}. However, experimenting on a virtual network does
not fit our research goal, which is to estimate the population of I2P peers
and assess the network's resistance to blockage.

We should note that throughout our study, we not only contribute additional
routing capacity to the I2P network, but also help in maintaining the distributed
network database. Considering only the main experiment over a period of three
months, each router under our control is configured to contribute
a shared bandwidth of 8 MB/s in each direction, with an observed maximum
usage of 5MB/s.

\section{Methodology}
\label{sec:methodology}

Since I2P is a distributed network without any centralized authorities, we
need to take a black-box approach to answer our research questions regarding
the size of the I2P network and its resistance to censorship.
In practice, there are several ways for an adversary to harvest I2P's network
database (netDb). For
instance, one can keep crawling the hard-coded reseed servers to fetch as many
\RI{}s as possible. However, to cope with such malicious activities,
reseed servers are designed so that they only provide the \emph{same} set of
\RI{}s if the requesting source is the same.
Nevertheless, an adversary
who has control over a large number of IP addresses can still continuously
harvest the netDb by crawling the reseed servers from different IP addresses.
Another way of harvesting netDb information
is to manipulate the netDb mechanism in an
aggressive manner through the DatabaseLookupMessage (DLM) interface.
Normally, peers that do not have a
sufficient amount of \RI{}s
in their netDb and peers that need to look
up \LS{}s will send a DLM to floodfill routers to request more
\RI{}s and \LS{}s.
Making use of this mechanism, adversaries could modify the source
code of the I2P router software to make their I2P clients repeatedly query floodfill
routers to aggressively gather more \RI{}s.

For the purposes of our research, the above approaches are impractical and
even unethical. Although one of the goals of this paper is to estimate the
population of I2P peers, which requires us to also collect as many
\RI{}s from the netDb as possible,
we need to conduct our study in a responsible manner. Our principle
is that experiments should not cause any unnecessary overheads or saturate any
resources of other I2P peers in the network. Liu et al.~\cite{Liu2014}
showed that crawling reseed severs only contributes 7.04\% to the total number
of peers they collected, while manipulating the netDb mechanism only
contributes 30.18\%.

Therefore, we choose an alternative method,
and opt to conduct our experiments
in a passive way by operating several routers that simply observe the
network. The primary goal of our experiments is to investigate
\emph{how many I2P routers one needs to operate and under what settings
to effectively monitor a significant portion of the I2P network with the least
effort}. In order to avoid the bandwidth
limitation of prior studies~\cite{Timpanaro2011}, all of our experiments are
conducted using dedicated private servers
instead of research infrastructure shared with other researchers.

\subsection{Machine Specifications}
\label{sec:SpecsExperiment}

\begin{figure}[t]
\centering
\includegraphics[width=1\columnwidth]{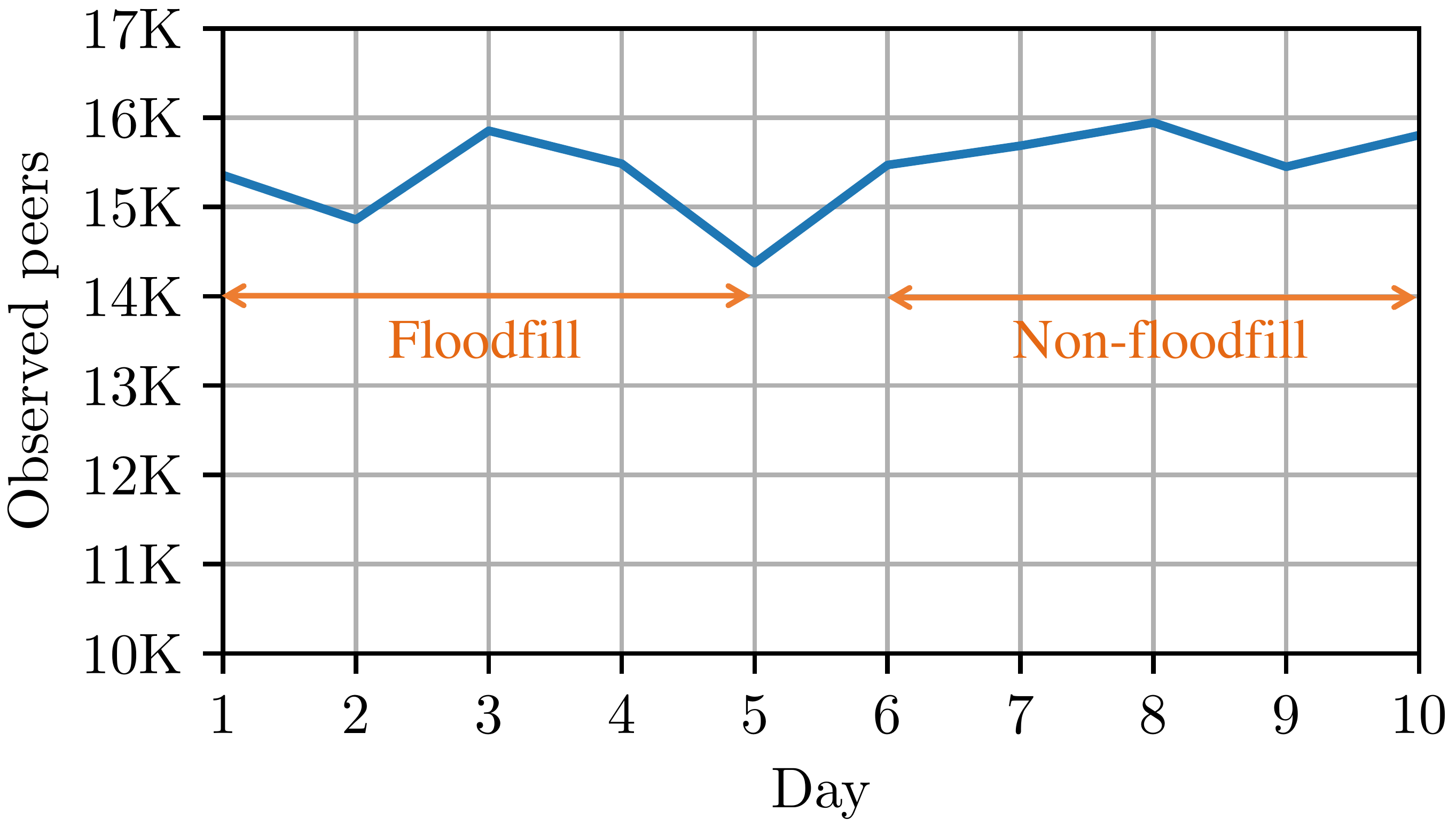}
\caption{Number of peers observed during our initial experiment for
assessing the impact of different hardware and software configurations.}
\label{fig:SpecsExperiment}
\end{figure}

Since there is no official guideline on how to operate a high-profile I2P
router, we employ a best-effort approach to determine what specifications are
sufficient to observe a significant amount of other I2P routers. Specifications
of interest include the hardware configuration of the hosting machine (e.g.,
CPU, RAM) and configuration parameters of the I2P router software (e.g., shared
network bandwidth, maximum number of participating tunnels, size of heap memory
for the Java virtual machine). Note that the official I2P router software is
written in Java. This is a necessary step in order to understand the I2P
software behavior. For example, increasing the number of connections allowed to
a router, without tuning the available Java heap space, can result in errors
that will force a router to restart. Similarly, if CPU is not adequate, a router
might drop connections, block, or increase latency. These are all situations
under which a router would be penalized by the I2P ranking algorithm and
therefore have less chances of being chosen to participate in peers' tunnels.
Consequently, a router that is not fine-tuned will have less visibility into the
I2P network than one that can maintain a high service quality. We empirically
investigate the upper bounds of a system's specifications to decide the resources
we will need to dedicate to our hosts.

Intuitively, we know that a higher-profile router will observe a larger number
of \RI{}s.
We first run an I2P router using a high-end machine with
a 10-core 2.40 GHz CPU and 16 GB of RAM.
The shared bandwidth of this router is then
set to 8 MB/s because the built-in bloom filter of the I2P router software is
limited to 8 MB/s. The maximum number of participating tunnels is set to 15K,
and 10 GB is allocated to the heap memory for the Java virtual machine. After
running this router for 10 days, five days in each mode (i.e., floodfill and
non-floodfill), we make the following observations:

\begin{itemize}

\item Total CPU usage always stays in the range of 4--5 Ghz.

\item Memory usage stays in the range of 3--4 GB most of the time.

\item The highest observed bandwidth usage is 5 MB/s.

\item The number of participating tunnels stays at around 4K,
while the highest observed number is approximately 5.5K tunnels.

\item All of the maximum values above are observed when operating
  in the non-floodfill mode.

\end{itemize}

As shown in Figure~\ref{fig:SpecsExperiment}, although the number of
peers observed during the non-floodfill mode is slightly higher than in the
floodfill mode, it constantly remains around 15--16K. Note that a peer is
defined by a unique hash value encapsulated in its \RI{}.
Based on these observations, we set up the (virtual)
machines used for our subsequent
experiments with the following upper-bound specifications:

\begin{itemize}

\item Three 2.4 GHz CPU cores totalling 7.2 GHz.

\item Five GB of RAM, four of which are allocated to the heap memory of the Java
  virtual machine and one for the rest of the system.

\item The maximum number of participating tunnels is set to 10K.

\item The maximum shared bandwidth is set to 8 MB/s, according to the
  maximum limit of the built-in bloom filter of the I2P router
software.

\end{itemize}

\subsection{Floodfill vs Non-floodfill Operation}
\label{sec:ModeSelect}

Although Figure~\ref{fig:SpecsExperiment} shows that the number of peers
observed in non-floodfill mode is slightly higher than in floodfill mode, it
is possible that this difference is the result of a fluctuation in the number
of daily peers during the study period. Therefore, we operated another 14
routers in both floodfill and non-floodfill mode simultaneously to prevent
any potential fluctuation in the number of daily peers from affecting
our observations.
These 14 routers are divided into two groups: non-floodfill and
floodfill, with seven routers in each group. For the routers in each group,
we gradually increase the shared bandwidth as follows:
128 KB/s, 256 KB/s, 1 MB/s, 2 MB/s, 3 MB/s, 4 MB/s,
and 5 MB/s. We pick 128 KB/s as the lowest bandwidth because it is
the minimum required value for a router to be able to gain the floodfill
flag~\cite{netDb}, while the highest value is based on the highest bandwidth
usage observed in our previous experiment (Section~\ref{sec:SpecsExperiment}).
We run these routers on machines with hardware specifications described earlier.

\begin{figure}[t]
\centering
\includegraphics[width=1\columnwidth]{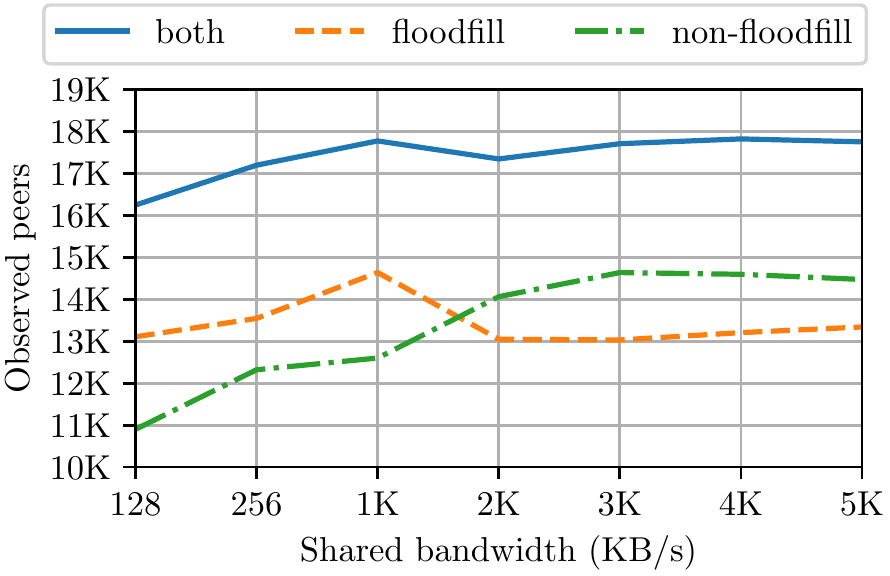}
\caption{Number of I2P peers observed when operating 14 nodes (7 in floodfill
  and 7 in non-floodfill mode) using an increasing amount of shared
bandwidth.}
\label{fig:ModeSelect}
\end{figure}

\begin{figure}[t]
\centering
\includegraphics[width=1\columnwidth]{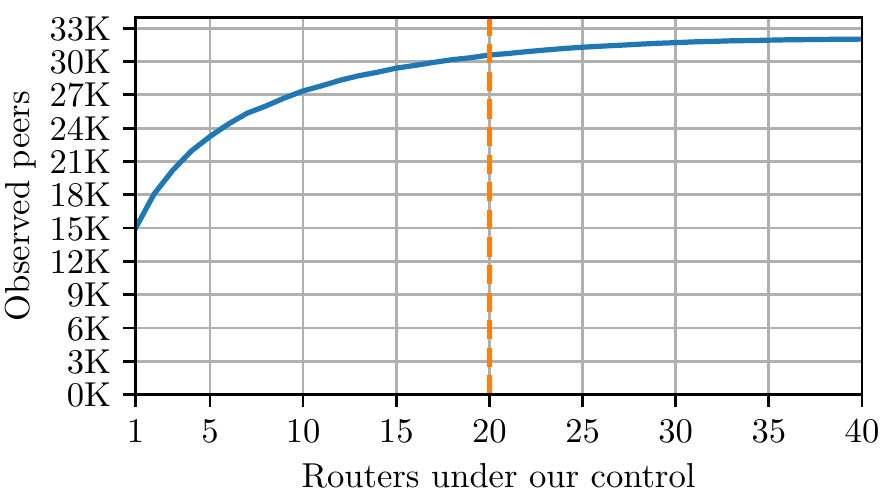}
\caption{Cumulative number of peers observed by operating 1--40 routers.}
\label{fig:NoRouters}
\end{figure}

Figure~\ref{fig:ModeSelect} shows that floodfill routers with shared bandwidth
lower than 2 MB/s observe 1.5--2K more peers than non-floodfill routers that
have the same shared bandwidth. On the other hand, non-floodfill routers with
shared bandwidth greater than 2 MB/s observe about 1--1.5K more peers than
floodfill routers of the same shared bandwidth. However, it is interesting that
when combining data from each pair of routers with the same shared bandwidth,
the total number of observed peers (upper line in the graph) stays at around
17--18K, regardless of the difference in shared bandwidth and the number of
observed peers in each mode. To explain this behavior, we first identify the
four primary ways I2P peers can learn about other peers in the network:

\begin{itemize}

\item As part of the bootstrapping process,
  a newly joined peer fetches \RI{}s
  from a set of hardcoded reseed servers
to learn a small portion of peers in the network.
Based on logs provided by the I2P router console, a
newly joined peer fetches around 150 \RI{}s from two reseed servers
(roughly 75 \RI{}s from each server).

\item A router that does not have enough \RI{}s
in its local storage
sends a DLM to floodfill routers to ask for more \RI{}s.

\item An active router is selected by other peers to
route traffic for them. This way, the router learns about other adjacent
routers in tunnels that it participates in. The higher the specifications a
router has, the higher the probability that it will be selected to participate
in more tunnels.

\item A floodfill router receives \RI{}s published by other ``nearby''
  non-floodfill routers or by other floodfill routers via the flooding
  mechanism. The ``nearby'' distance is calculated based on the \texttt{XOR}
  distance between the indexing key of two routers. The flooding mechanism is
  used when a floodfill router receives a DatabaseStoreMessage containing a
  valid \RI{} or \LS{} that is newer than the one previously stored in its local
  NetDb. In that case, the floodfill router ``floods'' the netDb entry to three
  others among its closest floodfill routers~\cite{netDb}.

\end{itemize}

We attribute the observed behavior to the last two of the above mechanisms,
as they are the main ways in which our routers
learn about other peers in the network. Since the two groups of routers used 
interact with the network in different ways,
each group obtains a particular
view of the network from a different angle, which the other group could not
observe. As a result, aggregating their data together gives us a better view
of the overall network. In summary, from this experiment we learn
that it is important to operate
routers in both non-floodfill and floodfill modes. By combining different 
viewpoints, we can gain a more complete view of the network.

\subsection{Number of Routers}
\label{sec:NoRouters}

Next, we investigate how many routers we need to run to observe a
significant part of the network. Prior to this work, Liu et al.~\cite{Liu2014}
used various methods to harvest the netDb:
crawling the reseed servers repeatedly, sending
DLM continuously to other floodfill routers, and running both floodfill and
non-floodfill routers. The authors claim the discovery
of 94.9\% of all routers in the network by comparing their collected data with
the \texttt{stats.i2p} statistic website~\cite{i2pstats}.
However, as we have confirmed with the I2P team,
the provided statistics cannot be considered as ground truth. This is because
the statistics are collected only from an average non-floodfill router
(i.e., not high bandwidth).
Furthermore, reported results are plotted using data collected over the last
thirty days, but not on a daily basis.
More recently, Gao et al.~\cite{Gao2017} operated
40 floodfill routers to collect \LS{}s and claimed the discovery of more
than 80\% of all ``hidden'' \ES{}s.
However, it is not clear which
hardware and software combination was used for operating those routers.
More importantly,
as we are interested in gathering \RI{}s but not \LS{}s,
operating all routers in a single mode (i.e., floodfill or non-floodfill) is not
ideal (see our discussion in Section~\ref{sec:ModeSelect}).

Therefore, we choose to run a total of 40 routers equally divided between both modes
(floodfill and non-floodfill). Each router is hosted on a
machine with the specifications defined in Section~\ref{sec:SpecsExperiment}. As
\RI{}s are written to disk by design so that they are available
after a restart~\cite{netDb}, we keep track of the netDb directory where these
records are stored. Note that although there is an expiration field in the
data structure of \RI{}, it is not currently used~\cite{routerAddress}. That
means the actual active time of a peer is unknown. In other words, the
existence of a given \RI{} only indicates the presence of the corresponding
peer in the network, but it does not provide an indication about until when a
peer was active.

Since floodfill routers apply a one-hour expiration time for all
\RI{}s stored locally, we choose to monitor the netDb directory on an
hourly basis to capture any new \RI{}. Every 24 hours
we clean up the netDb directory to make sure that we do not count
inactive peers on the next day. After running these routers for five days,
we calculate the cumulative number of peers observed daily across 40
routers.

Figure~\ref{fig:NoRouters} shows that operating 40 routers can help us
observe about 32K peers in the network. The number of observed peers
has a logarithmic relation to the number of routers under our control. The figure
also shows that the number of observed peers increases rapidly when increasing
the number of routers from one to 20,
and then increases slowly and converges to about 32K. In fact, the
aggregated number of observed peers from operating 20 routers already
gives us 95.5\% (i.e., more than 30.5K peers) of the total number of
observed peers.
Beyond 35 routers, each added router only contributes the observation of an extra
10--30 peers. Therefore, we conclude that 20 routers are sufficient
for obtaining a good view of the I2P network.

\section{Network Measurement}
\label{sec:NetworkMeasurement}

Taking the observations made in Section~\ref{sec:methodology} into
consideration, we conducted our measurements by operating
20 routers using the machine specifications defined
in Section~\ref{sec:SpecsExperiment}. These routers consist of 10 floodfill
and 10 non-floodfill routers. We collected \RI{}s observed by these
routers for a period of three months (from February to April, 2018).

\subsection{Population of I2P Peers}
\label{sec:i2pRouterPopulation}

\begin{figure}[t]
\centering
\includegraphics[width=1\columnwidth]{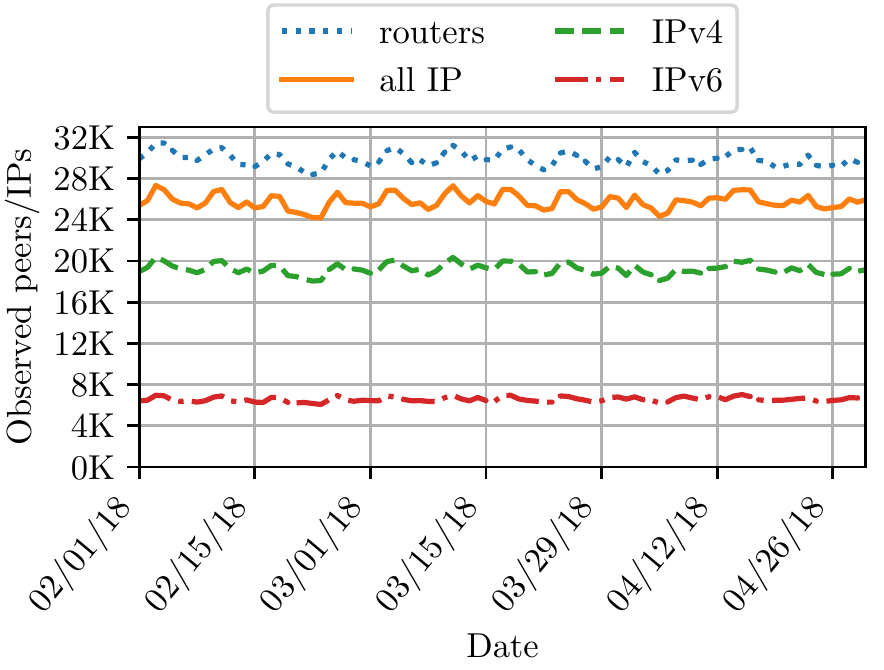}
\caption{Number of unique peers and IP addresses.}
\label{fig:i2pRouterPopulation}
\end{figure}

Figure~\ref{fig:i2pRouterPopulation} shows the number of unique I2P peers
and IP addresses, including both IPv4 and IPv6, observed during the
three-month period. The number of daily peers remains stable at around 30.5K. Note
that an I2P peer is identified by a cryptographic identifier, which is a
unique hash value encapsulated in its \RI{}. This identifier is generated
the first time the I2P router software is installed, and never changes
throughout its lifetime.

For the number of unique IP addresses, we count all unique IPv4 and IPv6
addresses (if supported by an I2P router) on a daily basis. Given that some peers
frequently change their IP address, as we discuss in
Section~\ref{sec:ipChange}, one would expect the total number of unique IP
addresses to be higher than the number of peers. However, as shown in
Figure~\ref{fig:i2pRouterPopulation}, the total number of IP addresses is
noticeably lower than the number of peers. By analyzing
the collected \RI{}s, we identified a large number of I2P peers whose
\RI{}s do not have a valid IP address field. In other words,
the public IP addresses of these peers are unknown.
We then analyzed other fields in the \RI{} of
these peers and discovered that there are two subgroups of peers within the
group of unknown-IP peers. These are firewalled and hidden peers. Firewalled
peers are operated behind NAT or strict firewall configurations. Hidden peers
only use other peers to route their traffic but do not help other peers to
route traffic since they do not publish their IP address in the network
database. By default, peers located in countries with poor Press Freedom
scores (i.e., greater than 50)~\cite{i2pBadCountries, worldPressFreedom} are
set to hidden. However, this setting can be modified to expose the peer to the
rest of the network to benefit a better integration, thus better performance.
We classify these two groups by examining the IP address field of
\emph{introducers} in each \RI{} file.

I2P provides a way for peers behind NAT or firewalls to communicate
with the rest of the network, using third-party introduction points (aka
\emph{introducers})~\cite{i2pIntroducers}. An I2P peer (e.g., Bob) who
resides behind a firewall that blocks unsolicited inbound packets, can choose
some peers in the network to become his introducers. Each of these introducers
creates an introduction tag for Bob. These tags are then made available to the
public as a way to communicate with Bob. Having Bob's public tags, another
peer (e.g., Alice) sends a request packet
to one of the introducers, asking it to introduce her to Bob. The introducer
then forwards the request to Bob by including Alice's public IP and port
number, and sends a response back to Alice, containing Bob's public IP and
port number. Once Bob receives Alice's information, he sends out a small
random packet to Alice's IP and port, thus punching a hole in his firewall for
Alice to communicate with him.

\begin{figure}[t]
\centering
\includegraphics[width=1\columnwidth]{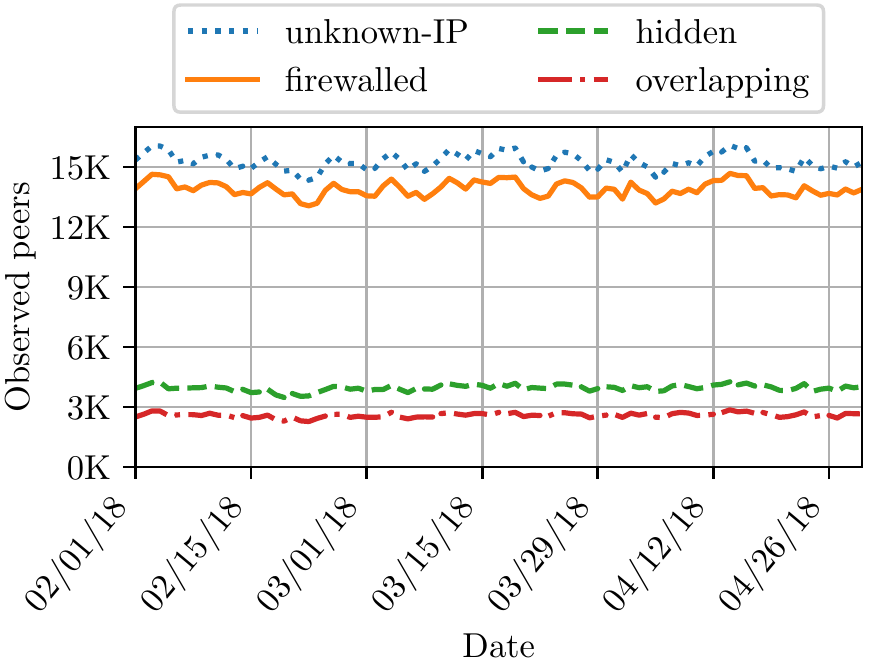}
\caption{Number of peers with unknown IP addresses.}
\label{fig:unknownRouters}
\end{figure}

By examining the IP address field of the introduction points in
\RI{}s, we can differentiate between firewalled and hidden
peers. A firewalled peer has information about its \emph{introducers} embedded
in the \RI{}, while a hidden peer does not.
Figure~\ref{fig:unknownRouters} shows the number of peers in each group. In
total, there are more than 15K unknown-IP peers per day, which consist of
roughly 14K firewalled peers and 4K hidden peers. Between these two
groups, there are about 2.6K overlapping peers. In other words, there are
2.6K I2P peers per day that have their status changing between firewalled
and hidden.

\subsection{Churn Rate}
\label{sec:churnRate}

I2P is a dynamic P2P network in which peers come and leave frequently. Prior
to this work, Timpanaro et al.~\cite{Timpanaro2015} conducted the first churn
study of I2P and reported the probability of an I2P peer going offline after
30 minutes to be around 15\%. However, the experiment was conducted for
only five days, and only eight floodfill routers were deployed. Liu et
al.~\cite{Liu2014} ran their experiment for around two weeks and reported that
19.03\% of the collected peers survived for one day, while 48.66\% of them
survived more than seven days. Overall, these works were
conducted over a short period of time and on a small scale, providing an
incomplete view of the churn rate of the I2P network. Moreover, none of the
previous studies mentioned the address changing phenomenon of peers in the
network, which often happens due to the fact that most ISPs do not usually
allocate a static IP address to residential Internet connections. In this
section, we analyze the collected \RI{}s to fill these research gaps.

\subsubsection{Peer Longevity}

\begin{figure}[t]
\centering
\includegraphics[width=1\columnwidth]{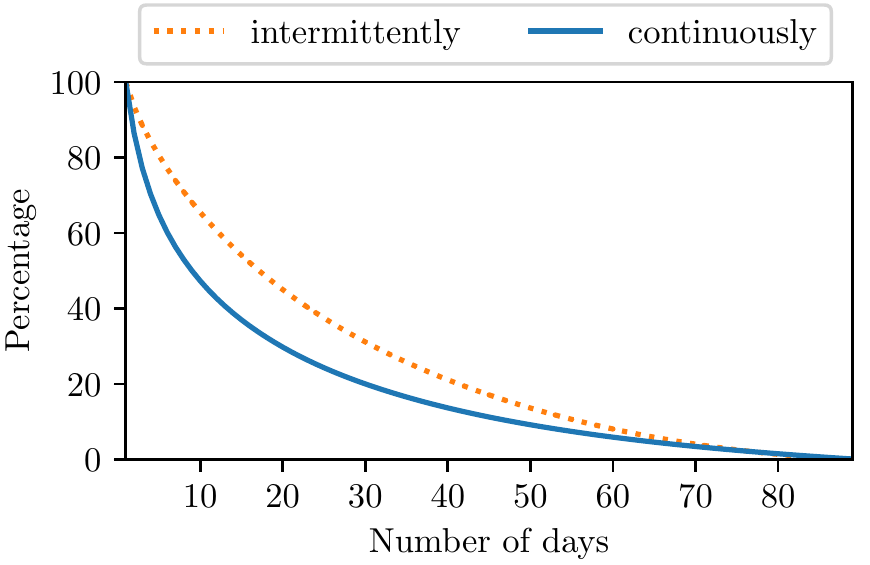}
\caption{Percentage of peers that we see in the network continuously or
intermittently for $n$ days.}
\label{fig:peerLongevity}
\end{figure}

Figure~\ref{fig:peerLongevity} illustrates the churn rate of I2P peers during
our three-month measurement. As shown in
Figure~\ref{fig:peerLongevity}, the percentages of peers staying in the network
for more than seven days are 56.36\% (continuously) and 73.93\%
(intermittently). That percentages of peers online longer than 30 days are
20.03\% (continuously) and 31.15\% (intermittently). Although I2P is a purely
distributed and dynamic P2P network, these results imply that
more than half of the peers stay stably in the network more than a week.
Compared with the churn rate of 48.66\% in 2014~\cite{Liu2014}, our findings
of both continuous and intermittent churn rates show that the network is
becoming more stable.

\subsubsection{IP Address Churn} 
\label{sec:ipChange}

\begin{figure}[t]
\centering
\includegraphics[width=1\columnwidth]{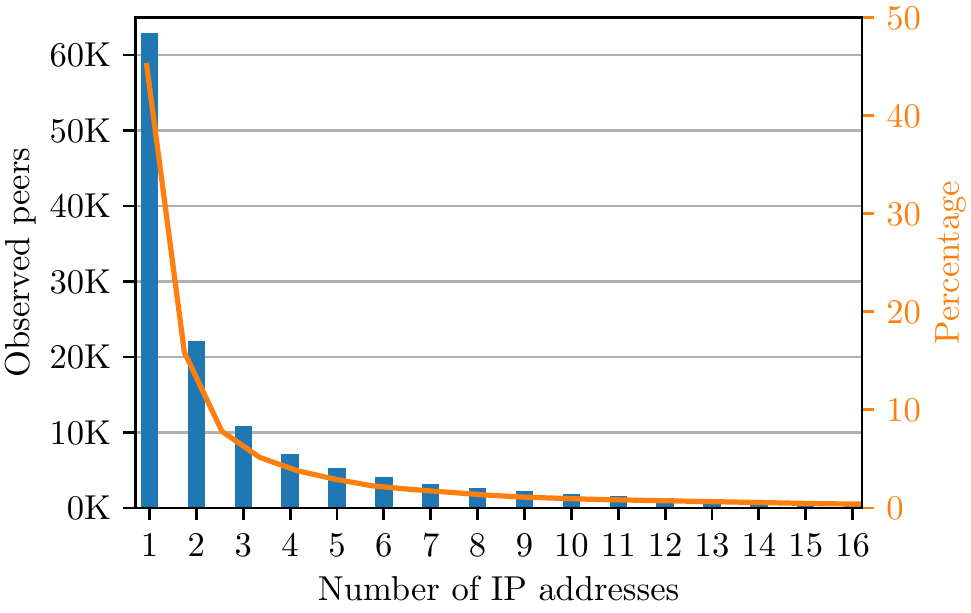}
\caption{Number of IP addresses I2P peers are associated with.}
\label{fig:ipChange}
\end{figure}

Since most ISPs do not allocate a static IP address for residential Internet
connections, it is common for peers to be associated with more than one IP
address. As shown in Figure~\ref{fig:ipChange}, there are 63K peers that are
associated with a single IP address (45\% of known-IP peers), while more
than 76K known-IP peers (55\%) are associated with at least two IP
addresses. Moreover, we notice a small group of 460 peers that are
associated with more than a hundred IP addresses during a period of three
months, occupying 0.65\% of the total number of known-IP peers. We
characterize this phenomenon in Section~\ref{sec:geographicDistribution}
when we study the geographic distribution of I2P peers.

\subsection{Peer Distribution}
\label{sec:netDistribution}

Peers in the I2P network are classified with different \emph{capacity flags} based
on their (1) operating mode (floodfill vs. non-floodfill), (2) reachability
(whether or not they are reachable by other peers), and (3) shared
bandwidth~\cite{netDb}. These capacity flags, denoted by a single capital
letter, are stored in the \RI{} file
of each peer. We are interested in understanding the percentage of each
peer type in the I2P network. Prior to this study,
Liu et al.~\cite{Liu2014} analyzed the
distribution of I2P peers across countries. However, the multiple IP
addresses phenomenon necessitates a more thorough approach for analyzing peers
that change address frequently. As mentioned in Section~\ref{sec:ipChange},
more than half of the known-IP peers are associated with two or more IP
addresses. In this section, we analyze two aspects of I2P peers: capacity
and geographic distribution.

\subsubsection{Peer Capacity Distribution}
\label{sec:routerCapDistribution}

Capacity flags are used by peers in the network for basic decisions, such as
peer selection for creating tunnels, and floodfill router selection for submitting
\RI{} and \LS{} information. The status of a peer is determined as follows:

\begin{itemize}

\item A floodfill router is denoted by an \emph{f} flag in its
capacity field, while a non-floodfill router does not have this flag.

\item The estimated shared bandwidth range of a peer is indicated by one of
  seven available letters: \texttt{K}, \texttt{L}, \texttt{M}, \texttt{N},
  \texttt{O}, \texttt{P}, and \texttt{X}, which correspond to less than 12KB/s,
  12--48 KB/s, 48--64 KB/s, 64--128 KB/s, 128-256 KB/s, 256-2000 KB/s, and more
  than 2000 KB/s, respectively.

\item The reachability of a peer is defined by \texttt{R} (reachable) or
  \texttt{U} (unreachable).

\end{itemize}

For example, the \texttt{OfR} flags found in the capacity field of a peer, mean
that the peer is a reachable floodfill router with a shared bandwidth of
128--256 KB/s. Analyzing these capacity flags provides us a better understanding
of peer capacity distribution in the network, and allows us to accurately
estimate the total amount of peers in the network.

\begin{figure}[t]
\centering
\includegraphics[width=1\columnwidth]{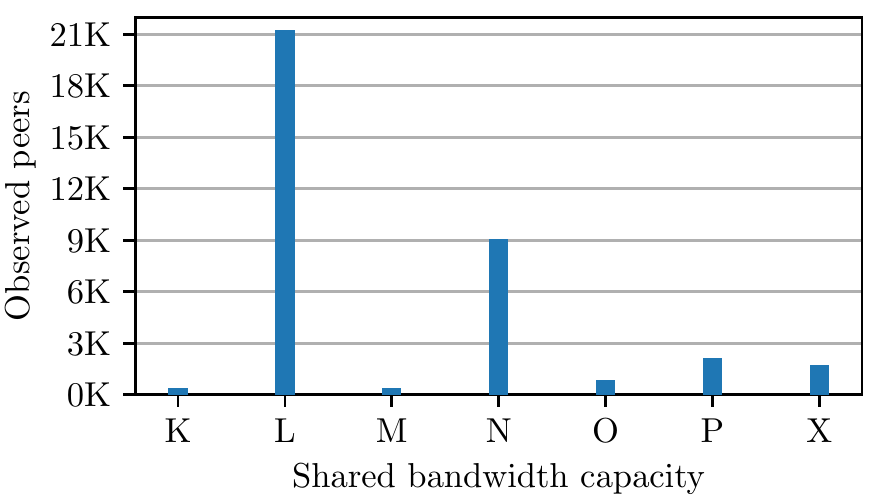}
\caption{Capacity distribution of I2P peers.}
\label{fig:routerCapDistribution}
\end{figure}

Our analysis in Figure~\ref{fig:routerCapDistribution} shows that
\texttt{L}-flagged peers are the most dominant in the network, with an average
of about 21K peers per day. This result complies with the fact that the
\texttt{L} flag is the default shared bandwidth of the I2P router software.
With more than 9K peers on a daily basis, \texttt{N}
is the second most dominant peer type. \texttt{P}, \texttt{X}, \texttt{O},
\texttt{M}, and \texttt{K} peers have an average of 2.1K, 1.8K, 875, 400, and
360 peers per day, respectively. In terms of operation mode, we observed an
average of 2.7K floodfill peers per day, which corresponds to 8.8\% of all peers
observed. Regarding peer reachability, the numbers of both \emph{reachable} and
\emph{unreachable} peers are almost the same most of the time, at around
15--16K. In other words, reachable and unreachable peers occupy roughly half of
the network each. Note that \emph{unreachable} peers include the unknown-IP
peers discussed in Section~\ref{sec:i2pRouterPopulation}.

We further analyze the bandwidth capacity distribution of each group: floodfill,
reachable, and unreachable. As shown in Table~\ref{table:routerCapDistribution},
while reachable and unreachable groups have a similar capacity distribution to
the whole network in which \texttt{L}-flagged type is the most dominant and
\texttt{N}-flagged type is the second, the floodfill group has the
\texttt{N}-flagged type as the most dominant, and the \texttt{L}-flagged type
comes second.

\begin{table}[t]
  \begin{center}
  \scalebox{0.9}{
  \begin{tabular}{l l r r r r}
    \multicolumn{1}{c}{\textbf{Bandwidth}} & & \multicolumn{1}{c}{\textbf{Floodfill}} & \multicolumn{1}{c}{\textbf{Reachable}} & \multicolumn{1}{c}{\textbf{Unreachable}} & \multicolumn{1}{c}{\textbf{Total}} \\
    \hline
    $ < 12$ KB/s     & \texttt{K}  & 0.10   & 1.14   & 1.27   & 1.18 \\
    12--48 KB/s      & \texttt{L}  & 26.82  & 66.62  & 75.81  & 69.67 \\
    48--64 KB/s      & \texttt{M}  & 2.16   & 1.44   & 1.24   & 1.31 \\
    64--128 KB/s     & \texttt{N}  & 62.06  & 36.79  & 26.08  & 29.74 \\
    128--256 KB/s    & \texttt{O}  & 5.18   & 3.15   & 2.88   & 2.87 \\
    256--2000 KB/s   & \texttt{P}  & 15.97  & 7.72   & 6.64   & 7.05 \\
    $ > 2000$ KB/s & \texttt{X}  & 13.76  & 6.44   & 5.49   & 5.76 \\
    \hline
  \end{tabular}
}
  \caption{Percentage of routers in different bandwidths, based on their
    floodfill, reachable, and unreachable status.}
  \label{table:routerCapDistribution}
  \end{center}
\end{table}

Note that the sum of all flags is not equal to 100\% for two reasons: (1) the
fluctuation in the bandwidth of a peer can frequently change its capacity flag,
and (2) for backwards compatibility with older software versions, a peer may
publish more than one bandwidth letter at the same time~\cite{netDb}. More
specifically, \texttt{P} and \texttt{X} flags are added since version 0.9.20,
and they override the previous highest bandwidth flag (\texttt{O} flag). In
order for older versions of the I2P router software to function normally, a peer
with a \texttt{P} or an \texttt{X} flag also has an \texttt{O} flag in its
capacity field.

Within the floodfill group, the total percentage of \texttt{P} and \texttt{X}
peers is around 30\%, greater than the percentage of \texttt{L}-flagged peers.
The result aligns with the fact that the floodfill mode is only enabled
automatically on peers that are configured with high bandwidth limits. The
current minimum requirement for a floodfill router is 128 KB/s of shared
bandwidth. With the current rules for automatic floodfill opt-in, a peer needs
to have at least an \texttt{N} flag in order to become a floodfill router
automatically~\cite{netDb}. However, Table~\ref{table:routerCapDistribution}
shows that there is a group of floodfill routers with \emph{lower} shared
bandwidth than required. This group includes \texttt{K}, \texttt{L}, and
\texttt{M}-flagged peers, which together comprise roughly 30\% of all floodfill
routers observed. This contradiction is due to the fact that operators can force
their routers to operate in floodfill mode by manually turning on this option in
the router console. As a consequence, the qualified floodfill routers are only
routers with a sufficient shared bandwidth to serve the netDb mechanism (i.e.,
\texttt{N}, \texttt{O}, \texttt{P}, and \texttt{X}-flagged routers).

Based on the above observation about floodfill routers, we deem those
\texttt{K}, \texttt{L}, and \texttt{M}-flagged floodfill routers to be manually
enabled and unqualified floodfill routers. We recompute the number of qualified
floodfill routers by combining the sets of \texttt{N}, \texttt{O}, \texttt{P},
\texttt{X} peers, and removing any peers that overlap with the sets of
\texttt{K}, \texttt{L}, \texttt{M} peers. Based on this calculation, 71\% of the
total floodfill routers observed are purely \texttt{N}, \texttt{O}, \texttt{P},
or \texttt{X}-flagged. Consequently, the number of qualified floodfill routers
should be $2700 \times 0.71 = 1,917$ routers. However, among these qualified
floodfill routers, there are also high-profile floodfill routers that are
manually enabled like ours. Therefore, the amount of floodfill routers that are
automatically enabled after meeting all of the ``health'' requirements must be
less than 1,917 routers, which matches the estimated number (i.e. around 1,700)
given on the official I2P website as of April, 2018~\cite{netDb}.

According to independent observations by I2P developers on the official I2P
website, approximately 6\% of the peers in the network are floodfill
routers~\cite{netDb}, but not 8.8\% as found above. We show that this difference
is the result of unqualified floodfill routers, which are manually enabled and
do not actually meet the minimum bandwidth requirements. Based on the percentage
of ``automatic'' floodfill routers in the network (i.e., 6\%), the population of
I2P peers is calculated as $1,917 \div 0.06 = 31,950$, approximately. This
result strengthens our hypothesis and observation from
Section~\ref{sec:NoRouters}, that running 40 routers allowed us to observe
around 32K peers in the network. Evidently, we can conclude with confidence that
using 20 routers one can monitor more than 95.5\% of the I2P network.

\subsubsection{Geographic Distribution}
\label{sec:geographicDistribution}

\begin{figure}[t!]
\centering
\includegraphics[width=1\columnwidth]{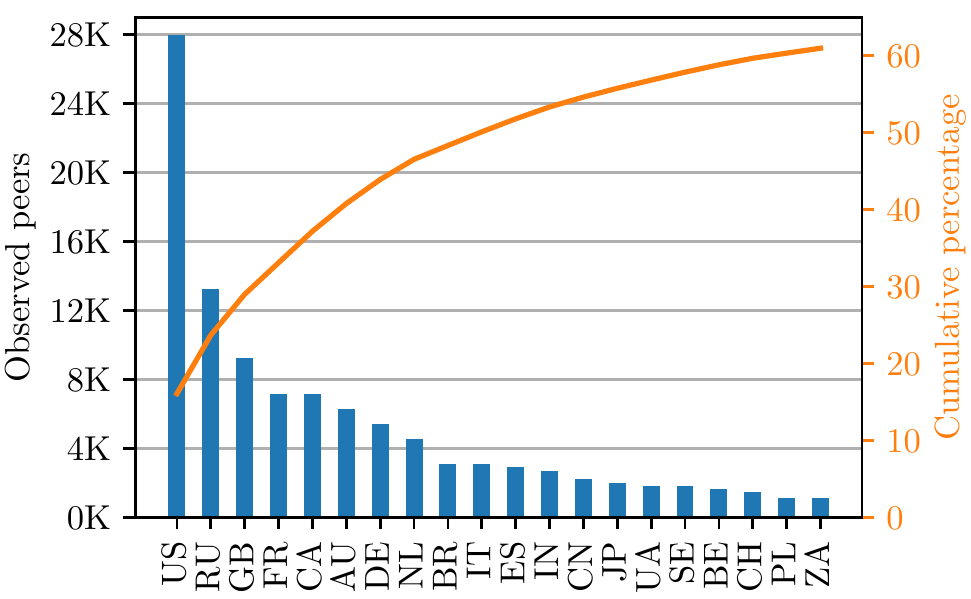}
\caption{Top 20 countries where I2P peers reside.}
\label{fig:country_ranking}
\end{figure}

Next, we utilize the MaxMind Database to map addresses of I2P peers to their
autonomous system number (ASN) and country. Since about half of the observed
peers are associated with more than one IP address, as discussed in
Section~\ref{sec:ipChange}, we need a proper way to count the
number of peers residing in each ASN/country. For each peer associated
with many IP addresses, we resolve these IP addresses into ASNs and countries
before counting them to avoid counting two different IP addresses belonging to
one peer. If two IP addresses of the same peer reside in the same
ASN/country, we count the peer only once. Otherwise, each different IP is
counted.

Figure~\ref{fig:country_ranking} shows the top 20 geographic locations of
I2P peers. United States, Russia, England, France, Canada, and Australia
occupy more than 40\% of peers in the network. The United States tops the
list with roughly 28K peers. Except for New Zealand, all Five Eyes
countries~\cite{fiveEyes} are in the top 10. This group of 20 countries
makes up more than 60\% of the total number of peers observed, while the
rest is made up of peers from 205 other countries and regions. Among 32
countries with poor Press Freedom scores (i.e. greater than
50)~\cite{worldPressFreedom}, there are 30 countries with a combined total
of 6K I2P peers. China leads the group with more than 2K
peers. Singapore and Turkey follow with about 700 and 600 peers observed
in the network, respectively.

Since China actively blocks access to Tor~\cite{Ensafi2015, Winter2012} and
VPN~\cite{Bloomberg2017,Reuters2017}, a portion of Chinese users seem to use
the I2P network instead. The number of Chinese users may be expected to
increase if more \emph{outproxies} become steadily available in the network.
Although China is one of the countries where I2P peers are configured to be in
hidden mode by default~\cite{i2pBadCountries, worldPressFreedom}, a router
operator can always turns off this setting to make his router more reachable,
thus improving performance.

\begin{figure}[t!]
\centering
\includegraphics[width=1\columnwidth]{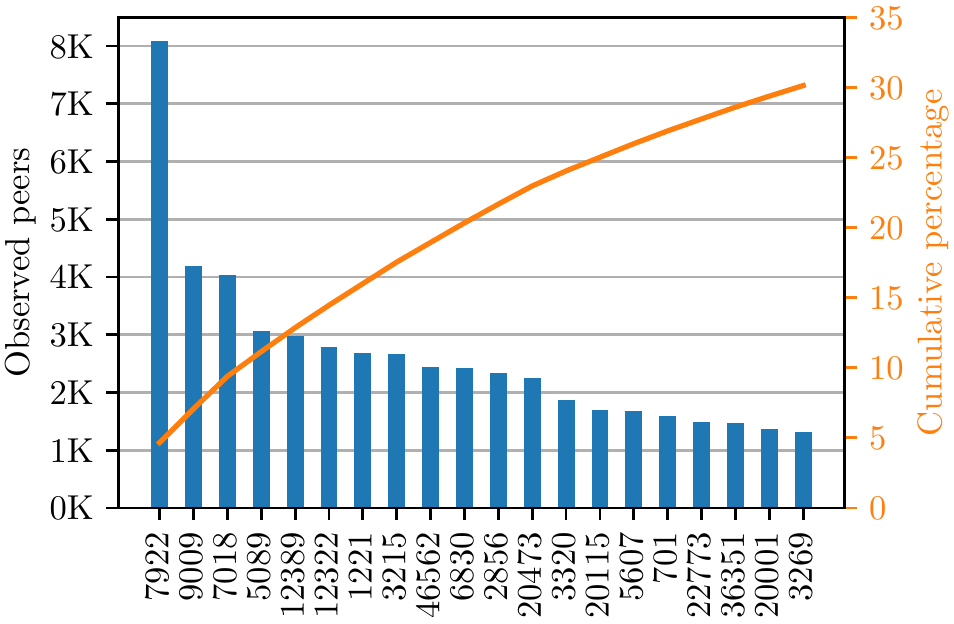}
\caption{Top 20 autonomous systems where I2P peers reside (the x axis
corresponds to the AS number).}
\label{fig:asn_ranking}
\end{figure}

Figure~\ref{fig:asn_ranking} shows 20 autonomous systems from which most
addresses originate. AS7922 (Comcast Cable Communications, LLC) leads the list
with more than 8K peers. Together these 20 ASes make up more than 30\%
of the total number of peers observed.

\begin{figure}[t!]
\centering
\includegraphics[width=1\columnwidth]{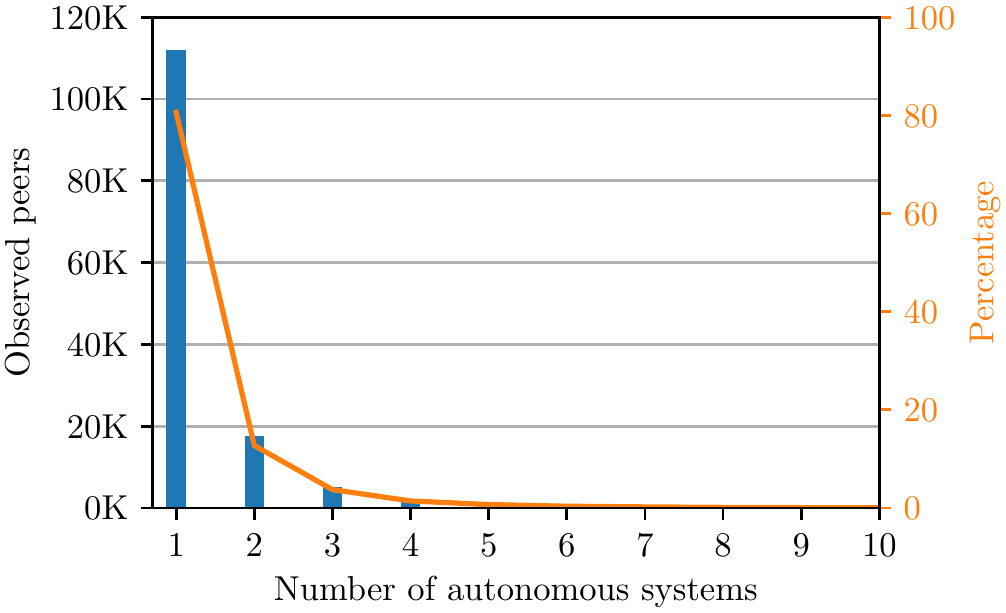}
\caption{Number of autonomous systems in which multiple-IP peers reside.}
\label{fig:asnChange}
\end{figure}

As mentioned in Section~\ref{sec:ipChange}, 58.9\% of peers change their
address at least once. We are also interested in analyzing this change in
terms of the geographic distribution of these peers. By mapping their IP
addresses to ASN and country, we find that most peers stay in the same
autonomous system or the same geographic region in spite of their association
with multiple IP addresses. This observation is reasonable given that although
ISPs frequently rotate different IP addresses dynamically for residential
Internet connections, these addresses often belong to the same subnet.
However, we notice a small portion of peers changing their IP addresses
repeatedly between different autonomous systems and countries. The
highest number of autonomous systems that a peer associates with is 39,
while the highest number of countries in which a peer resides in is 25.
Figure~\ref{fig:asnChange} shows the number of autonomous systems in which I2P
peers reside in. More than 80\% of peers only associate with one ASN,
while 8.4\% of peers are associated with more than ten different ASes.
Based on a discussion with one of the I2P developers, one of the possible
reasons for this phenomenon is that some I2P routers could be operated behind
VPN or Tor servers, thus being associated with multiple autonomous systems. Note
that users of Tails~\cite{tails} (until version 2.11) could use I2P over Tor
as one of the supported features.

A limitation of using MaxMind is that when mapping IP addresses to ASNs and
countries, there are around 2K addresses that we could not resolve using this
dataset. Nonetheless, this does not mean that we could not identify
2K peers. Our results in Section~\ref{sec:ipChange} show that more than 55\%
of known-IP peers are associated with more than one IP address. Therefore, the
actual number of peers whose ASN and country we could not identify are just
those peers that are associated with only one IP address we could not resolve.
As mentioned in our discussion of ethical considerations, we do not use any of
the more accurate public APIs on the Internet to resolve these IP addresses
for privacy reasons.

\section{Censorship Resistance}
\label{sec:blockingResistance} 

Due to the centralized network architecture of Tor, it is relatively easy for
a censor to find and block all public Tor routers. To cope with this blocking
susceptibility, several studies have aimed to enhance the
blocking resistance of Tor~\cite{Matic2017, Zamani2017, Ensafi2015,
Winter2012}. Despite its decentralized design, I2P is also susceptible to
censorship, but, to the best of our knowledge, its resistance to censorship
has not been extensively studied---we focus on this aspect in this section.

\subsection{Reseed Server Blocking}
\label{sec:reseedBlocking} 

Knowing the bootstrapping mechanism of I2P, a censor can easily block access
to the reseed servers to disable the I2P bootstrapping process. As a
consequence, reseed servers present a single point of blockage,
similarly to Tor's directory servers
(e.g., as was the case when they were blocked from China in 2009~\cite{ChinaTorBlock}).
Given the current design of I2P, a new
peer cannot connect to the rest of the network if it cannot bootstrap via
some reseed servers.

In April 2017, there was a post on the I2P developer forum reporting that
reseed servers were blocked in China~\cite{ChinaBlockade}. We attempted to test
the reachability of hardcoded reseed servers from some of our vantage points
hosted inside China and found that some of them were still accessible.
Moreover, the analysis in Section~\ref{sec:netDistribution} shows that China
is among the top-20 countries where most I2P peers reside. A previous
study~\cite{ensafi2015analyzing} shows two possibilities for our observation.
First, the report could be a case of small-scale blocking conducted at provincial
ISPs, but not a uniform nationwide blockage. Second, the Great Firewall 
of China (GFW) sometimes fails to block access to destinations that it normally
blocks. It is worth noting that the current I2P network can only be used as a
self-contained network most of the time due to the intermittent availability
of \emph{outproxies}. In addition, because the network is still small, it probably
has not yet become a target of censorship by the GFW. However, once the
network grows larger with more stable support of \emph{outproxies} to the Internet,
large-scale blocking is unavoidable.

The I2P developers have foreseen a situation in which all reseed servers
are blocked. Thus, a built-in function of the I2P router software is provided
to allow for manual reseeding. With this feature, every active I2P peer can
become a manual reseeder. Specifically, the function can be used to create a
reseed file called \texttt{i2pseeds.su3}. The file can then be shared with other
peers that do not have access to any reseed servers for the bootstrapping
process. The sharing can be done via a secondary channel,
similar to how Tor distributes bridge nodes (e.g., emails, file-sharing
services). Under this circumstance, a censor who wants to prevent local users
from accessing I2P has to find and block all addresses of active I2P peers.
However, since I2P is a distributed P2P network, it is challenging to obtain a
complete view of the whole network. We investigate the effectiveness and the
efficiency of this blocking approach next.

\subsection{Probabilistic Address-Based Blocking}
\label{sec:probabilisticBlocking}

We begin by considering a censor who 
tries to monitor the network and gather information about active peers 
(i.e., IP address and port), thus being able to prevent local
users from accessing the network. We then evaluate the blocking resistance of
an I2P peer and the usability of the I2P network under aggressive blocking
pressure.

\subsubsection{Setting} 

The probabilistic blocking model comprises (1) a group of monitoring routers
operated by a censor (e.g., ISP, government) and (2) a victim whom the censor
wants to prevent from accessing I2P. By operating some routers in the network,
the censor can acquire information about a large portion of potential peers
that the victim may need to contact in order to access the network, thus being
able to prevent the victim from accessing the network. The blocking rate is
then calculated by the rate of peer IP addresses seen in the netDb of the
victim, which can also be found in the netDb of routers that are controlled by
the censor.

\subsubsection{Blocking Resistance Assessment} 

We consider a long-term I2P node
who has been participating in the network and has many \RI{}s
in its netDb, which is about to be blocked.
To simulate the censor, we use IP addresses of daily active
peers observed by 20 routers under our control. For the victim, we run
an independent router outside the network that we use to host our
20 routers.

\begin{figure}[t]
\centering
\includegraphics[width=1\columnwidth]{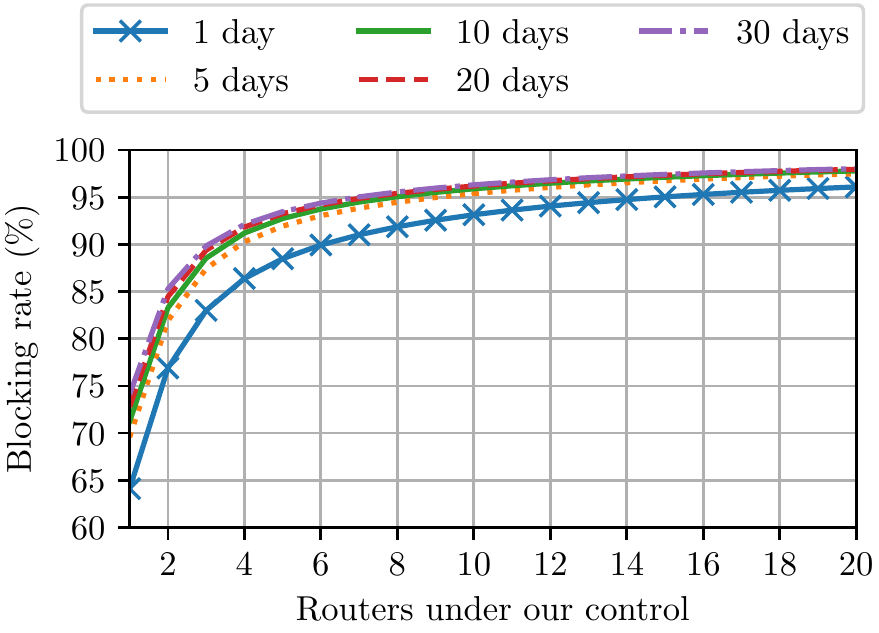}
\caption{Blocking rates under different blacklist time windows.}
\label{fig:probabilisticBlocking}
\end{figure}

The blue line (lowest) in Figure~\ref{fig:probabilisticBlocking} shows the cumulative
successful blocking rate of an adversary obtained by running 1--20 routers for
one day.
By operating 20 routers in the network, a censor can block more than 95\%
of peer IP addresses known by the victim, while 90\% can be blocked with just
six routers.

The above blocking rate is calculated based on the assumption that the censor
only uses IP addresses collected on a single given day. However, the actual situation
could be even worse. Previous studies on Tor have shown that once an IP
address is found to be joining an anonymous communication network or participating in
other types of network relays (e.g., VPN servers),
it may get blacklisted for several days, and sometimes even for more than a
month~\cite{Fifield2016, Singh2017}. We utilize the results obtained from the
churn rate analysis in Section~\ref{sec:churnRate} to examine how blocking
can be more severe if the IP blacklist time window expands to a
period of 5, 10, 20, or 30 days.

We find that if the censor expands the blacklist time window from one to
five days, the blocking rate increases to more than 97\% with 20 routers,
or 95\% with only 10 routers. Moreover, if the blacklist time windows expands
to a period of 10, 20, and 30 days, the blocking rates increase to
above 98\% with 20 routers, and about 96\% with only 10 routers.

As shown in Figure~\ref{fig:probabilisticBlocking}, five days would be
sufficient to achieve a high blocking rate. This is within the capabilities
offered by high-end firewalls used for nationwide censorship,
which can easily keep such a large number of rules.

\subsubsection{Network Usability Evaluation}

Since the address-based blocking implemented in the GFW of China uses the null
routing technique to route unwanted packets to a black-hole router, we
configure our upstream router to silently drop all packets that contain peer IP
addresses that we observed from the I2P network. We then set up three testing
\ES{}s to test the impact of the address-based blocking to the page load
time. These \ES{}s are designed with a simple and small html file to avoid
wasting bandwidth of the overall network. In addition, we conduct the test on
our own \ES{}s instead of publicly known \ES{}s to make sure that our
experiment does not disrupt legitimate users of those \ES{}s. We first crawl
our \ES{}s to test their average normal load time. The result in
Figure~\ref{fig:probabilisticBlocking} shows that a censor can block about
65\% to 98\% of peer addresses found in a victim's netDb. We then crawl these
\ES{}s 10 times for each blocking rate applied, measure the page load time,
and count the number of timed out requests (i.e., an HTTP 504 is returned).

\begin{figure}[t]
\centering
\includegraphics[width=1\columnwidth]{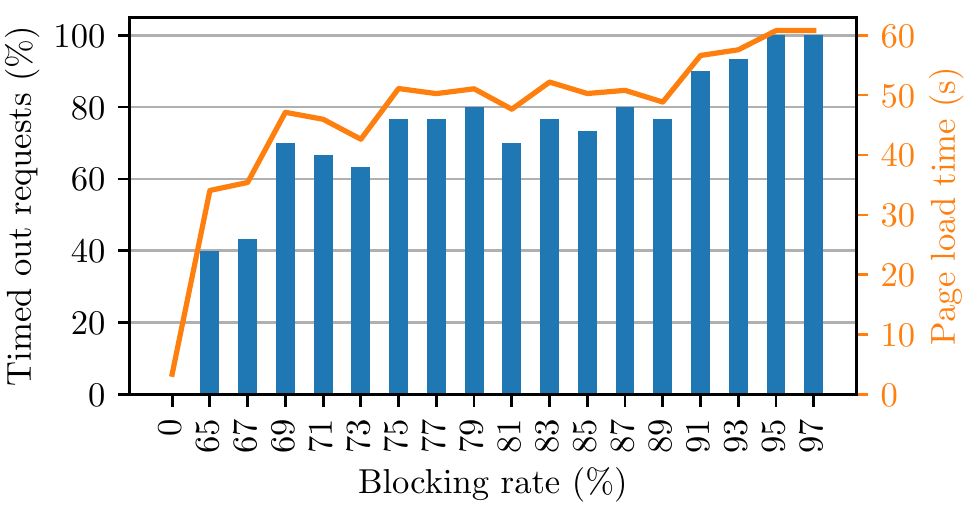}
\caption{Percentage of timeout requests and page load latency in the presence
of blockage.}
\label{fig:blockEval}
\end{figure}

Figure~\ref{fig:blockEval} shows that the average load time of our \ES{}s is
3.4 seconds without blockage. By blocking other peers with a rate of
65\%, a censor could already introduce a latency of more than 20 seconds to
the page load time and make 40\% of requests timed out. Any blocking rates in
the range of 70--90\% could cause a significantly higher latency in page load
time (i.e., more than 40 seconds), with the number of timed out requests
occupying more than 60\% of total requests. Blocking rates higher than 90\%
heavily depreciate the usability of the network, with 95--100\% of requests
timed out.

\section{Discussion}
\label{sec:discussion}

\subsection{Potential Solutions to Blocking}

Since more and more oppressive regimes attempt to prevent local users from
accessing the Tor network, Tor provides users in such restricted regions with
a set of special relays called \emph{bridges}~\cite{torBridges}. Similarly,
I2P can adopt the use of bridges to help those restricted users to
access the network, along with a non-fingerprintable traffic pattern currently
in development~\cite{i2pNTCP2}. While the Tor community may have a difficult
time recruiting bridges because new bridges are often found and blocked
quickly~\cite{Ensafi2015}, I2P has a higher potential to adopt the use of
bridges because of the high churn rate of its dynamic and decentralized
network.

Despite the high blocking rates shown in
Section~\ref{sec:probabilisticBlocking}, we notice a portion of peer IP
addresses could not be blocked. These IP addresses often belong to newly joined
peers. Therefore, a potential solution is to use these peers as bridges for
restricted users. Since these peers are newly joined, they are less likely
discovered and blocked immediately by the censor.

Utilizing newly joined peers as bridges, however, may only be suitable for
censored users who need to access I2P for a short period of time. If the peers
stay in the network long enough, they will be discovered by the monitoring
routers of the censor and eventually will be blocked. A potential approach to
remedy this problem is to use newly joined peers in combination with the
firewalled peers discovered in Sections~\ref{sec:i2pRouterPopulation} for a
more sustainable censorship circumvention. 

According to Figure~\ref{fig:unknownRouters}, there are around 14K
firewalled peers in the network on a daily basis. Without a public IP
address, the censor cannot apply the address-based blocking technique
introduced in Section~\ref{sec:probabilisticBlocking}. In the current I2P
design, the chance that a censor can discover the IP address of these
firewalled peers depends on the probabilities that the routers under the
censor's control (1) are selected to be introducers for these peers, and (2) 
they directly interact with these firewalled peers.

Except for implementing an infrastructure to collect and distribute bridges,
no overhead is introduced to any parties in the aforementioned solution. Since
most active peers in the network are selected to help other peers to route
traffic by default, the above approaches only changes how censored peers pick
non-blocked peers in order to access the rest of the network. Consequently,
utilizing newly joined peers in combination with firewalled peers can be a
potentially sustainable solution for restricted users who need longer access
to the network.

\subsection{From Blocking to Other Type of Attacks}

Although this study focuses on the problem of blocking access to I2P, the
probabilistic blocking model we introduced is not simply an effort to block
access to the I2P network. If a censor cannot completely prevent a local user
from accessing the network, it can conduct attacks such as traffic analysis to
deanonymize that user (e.g., revealing which destination is being visited by the
user).

For instance, after blocking more than 95\% of active peers in the network, the
attacker can inject malicious routers. He then configures the local network
firewall in a fashion that forces the victim to use the attacker's routers to
connect with the rest of the I2P network. In this case, the victim is
bootstrapped into the attacker's network. The attacker can facilitate this
process by whitelisting the group of malicious routers under their control,
while repeatedly blocking addresses of other active peers. By narrowing down the
victim's view of the network, the attacker is a step closer to conducting
several types of attacks, including the deanonymization attack mentioned
above~\cite{i2pThreat, Herrmann2011}.

\section{Conclusion}
\label{sec:conclusion}

In this work, we conducted a measurement study to
better understand the I2P anonymity network, which then allowed us to examine
its censorship resistance. Although I2P is not as popular as Tor, mainly
because it is used as a self-contained anonymity network,
the results of our measurements show that the network size is consistent over
the three-month study period, with roughly 32K daily active peers in the
network. Among these peers, about 14K of them are connecting to the I2P
network from behind NAT/firewall. During our three-month
study, we also discover a group of about 6K peers from countries with poor
Press Freedom scores.

We show that a censor can easily prevent local users from accessing the I2P
network at a relatively low cost, despite its decentralized nature.
Although the victim in our censorship resistance evaluation is
assumed to be a long-term and \emph{strong} peer that has been uninterruptedly
participating in the network, we show that a censor can still block more than
95\% of peer IP addresses found in the victim's netDb. This blocking rate can
be achieved by operating only 10 routers in the network, while applying
different blacklist time windows and running more routers (e.g., 20 routers)
can help the censor to achieve a blocking rate of almost 100\%.

As part of our
future work, we plan to expand our research by studying the feasibility of
using newly joined peers in combination with firewalled peers as bridges for
those peers that are blocked from accessing the network.

\section*{Acknowledgments}

We would like to thank our shepherd,
Mirja K\"uhlewind, the anonymous reviewers, and
the following members of the I2P team for their valuable
feedback: Sadie Doreen, str4d, echelon, meeh, psi, slumlord, and zzz.

\bibliographystyle{ACM-Reference-Format}
\bibliography{main}

\end{document}